\documentclass{article}
\usepackage[margin=3cm]{geometry}
\usepackage{authblk}

\usepackage{upgreek}

\usepackage{amsmath,amssymb,amsfonts}
\usepackage[nocompress]{cite}
\usepackage{caption} %%% to align caption left
\usepackage{makecell} %To write two lines cell in table
\usepackage[shortlabels]{enumitem}

\usepackage{ulem} %For strikethrough test
\numberwithin{equation}{section}

\usepackage{hyperref}
\hypersetup{colorlinks=true,allcolors=blue,urlcolor=cyan}

\title{Non-minimal fluid Lagrangian couplings}

\author[1,2]{Christian G B\"ohmer\footnote{Email: c.boehmer@ucl.ac.uk}}
\author[1]{Erik Jensko\footnote{Email: erik.jensko@ucl.ac.uk}}
\author[1,3]{Eissa Al-Nasrallah\footnote{Email: eissa.alnasrallah.24@ucl.ac.uk}}
\affil[1]{Department of Mathematics, University College London, \authorcr 
Gower Street, London WC1E 6BT, UK\medskip}
\affil[2]{Astrophysics Research Centre, School of Mathematics, \authorcr 
Statistics and Computer Science, University of KwaZulu-Natal, \authorcr 
Private Bag X54001, Durban 4000, South Africa\medskip}
\affil[3]{Department of Physics, College of Science, Kuwait University, \authorcr 
Sabah Al Salem University City, P.O. Box 5969, Safat 13060, Shadadiya, Kuwait}

\date{\today} 

\begin{document}

\renewcommand{\arraystretch}{1.2} % makes the tables prettier
\setlength{\tabcolsep}{1ex} % makes the tables prettier
\setlength{\extrarowheight}{1ex} % makes the tables prettier

\maketitle

\begin{abstract}
Gravitational models with non-minimal couplings involving functions of the matter Lagrangian and curvature have become popular in recent decades. By coupling the matter Lagrangian directly to the gravitational Lagrangian, one hopes to construct theories that can explain dark energy or dark matter without introducing additional sources. When this matter Lagrangian describes a perfect fluid, some technicalities are involved in its variational formulation. We present a careful derivation of the gravitational field equations together with the complete set of fluid equations using two approaches: Schutz's velocity potential and Brown's Lagrange multipliers. We find that with non-minimal couplings, the energy-momentum conservation equations give rise to a non-vanishing contribution which acts in a different direction depending on the approach. This also leads to the emergence of different effective thermodynamic quantities such as the number density, chemical potential, free energy, and temperature. We demonstrate the non-equivalence of the Lagrangian formulations of Schutz and Brown for these types of models and provide a detailed interpretation of our results. 
\end{abstract}

\clearpage 

\tableofcontents

\clearpage

\section{Introduction}
\label{introduction}

The observed accelerated expansion of the universe has inspired the development of modified theories of gravity in various forms. Many approaches have been taken in the literature, largely modifying the Einstein-Hilbert action, such as with additional fields or by considering higher-order curvature invariants~\cite{Clifton:2011jh,CANTATA:2021asi,Copeland:2006wr}. In particular, modifications with interactions between dark matter and dark energy have seen recent attention in relation to the cosmic tensions~\cite{Wang:2016lxa,Wang:2024vmw,CosmoVerseNetwork:2025alb}. Such interactions can be realised via non-minimal couplings, often in the context of scalar field dark energy \cite{Amendola:1999er,Pourtsidou:2013nha}.
An alternative geometric way to model dark energy was proposed in~\cite{NOJIRI2004137}, by introducing couplings between curvature and the matter Lagrangian $L_\textrm{m}$.
A distinctive feature of non-minimal curvature-matter couplings is that the matter energy-momentum tensor is generally not conserved; depending on the matter Lagrangian, this non-conservation may induce non-geodesic motion through an additional force term~\cite{Bertolami:2007gv,Bertolami:2008ab}.
%A distinctive feature of non-minimal curvature-matter couplings is their induced non-geodesic motion, where the energy-momentum conservation equation acquires an additional force term~\cite{Bertolami:2007gv,Bertolami:2008ab}.
A natural generalisation of these non-minimal matter couplings was proposed in~\cite{Harko:2010mv} by considering general Lagrangians of the form $f(R,L_{\rm m})$, which can be seen as an extension to the well-known $f(R)$ gravity models~\cite{Sotiriou:2008rp,DeFelice:2010aj}. In this paper, we revisit the non-minimal matter-curvature coupling models using the relativistic fluids approach developed by  Schutz~\cite{Schutz:1970} and  Brown~\cite{Brown_1993}.

For cosmological applications, matter is often modelled as a relativistic perfect fluid \cite{Misner:1973prb}. The energy-momentum tensor for such a fluid is given by
\begin{align}
\label{Fluid_T}
    T^{\mu\nu} = \bigl(\rho + p \bigr)U^\mu U^\nu + p g^{\mu\nu} = \rho U^\mu U^\nu + ph^{\mu\nu} \,,
\end{align}
where $\rho$ is the fluid's density, $p$ is the fluid's pressure, $U^\nu$ is the unit 4-velocity of the fluid satisfying $U^\nu U_\nu = -1$, and $h^{\mu\nu}=U^\mu U^\nu+g^{\mu\nu}$ is the projection tensor perpendicular to the 4-velocity. It is well-known that the Lagrangian generating~\eqref{Fluid_T} is not unique, with $L_{\rm m} = p$ or $L_{\rm m} = -\rho$ being the most common examples. 

It was shown in~\cite{Bertolami_2008} that when considering non-minimal curvature-matter couplings, different Lagrangians lead to different field equations and force terms in the conservation equations. A fully consistent approach was taken to derive the field equations and the conservation equation for different fluid approaches in~\cite{Boehmer:2025afy} for the $f(R,T)$ model, where $T$ is the trace of the energy-momentum tensor~(\ref{Fluid_T}). In this paper, we follow the same `first principles' approach for $f(R,L_{\rm m})$ models, where we consider the velocity potential approach for relativistic fluids proposed by Schutz~\cite{Schutz:1970}, and the Lagrange multipliers approach proposed by Brown~\cite{Brown_1993}. Notably, similar to the previous work~\cite{Boehmer:2025afy}, the two approaches lead to different equations of motion and different thermodynamic interpretations. 

A general action for the proposed model is given by
\begin{align}
\label{general_action}
    S = \frac{1}{2\kappa} \int f(R,L_{\rm m})\sqrt{-g}\, d^4x + \int \mathcal{L}_{\rm m} d^4x + 
    \int \mathcal{C}_{\rm m} d^4x\,,
\end{align}
where $\mathcal{L}_{\rm m}$ is the matter Lagrangian density and $L_{\rm m}$ is the matter Lagrangian, related by $\mathcal{L}_{\rm m}= L_{\rm m}\sqrt{-g}$. For completeness, we include the constraint density $\mathcal{C}_{\rm m}$ which will be needed to ensure that all matter equations of motion are appropriately taken into account. When dealing with perfect fluids using Brown's approach, for example, the constraints enforce the conservation of particle number and entropy flux along the fluid flow, and fix the flow lines on the spacetime boundary. On the other hand, the constraint Lagrangian density is zero $\mathcal{C}_{\rm m} = 0$ in Schutz's approach, highlighting the different mathematical structure of both frameworks.

In both approaches, it is important to pay attention to which of $p$ or $\rho$ is the fundamental quantity and which is derived. Both quantities are related through the first law of thermodynamics, given by the Helmholtz relations
\begin{align}
    dp &=n\,d\upmu -n\mathcal{T}\,ds 
    \label{thermodynamic_Schutz}\,, \\
    d\rho &=\upmu\,dn +n\mathcal{T}\,ds 
    \label{thermodynamic_brown}\,.
\end{align}
Here, $n$ is the particle number density, $\upmu$ is the chemical potential, $\mathcal{T}$ is the temperature, and $s$ is the entropy per particle. The chemical potential is defined by
\begin{align}
\label{chemical_potential}
    \upmu = \frac{\rho + p}{n} \,.
\end{align}
In Schutz's approach, the fundamental thermodynamic quantity is the pressure $p(\upmu,s)$, and the energy density is derived from the pressure 
\begin{align}
\label{density_relation_Schutz}
     \rho(\upmu,s) = \upmu\frac{\partial p}{\partial\upmu} -p \,.
\end{align}
It is the converse in Brown's approach, where the energy density $\rho(n,s)$ is the fundamental quantity and the pressure is the derived quantity
\begin{align}
\label{pressure_relation_Brown}
    p(n,s)=n\frac{\partial\rho}{\partial n}-\rho\,.
\end{align}
We see that the thermodynamic variables also change in each approach: in Schutz's approach, the pressure depends on the chemical potential $\upmu$ and entropy per particle $s$, whereas in Brown's approach, the energy density depends on the particle number density $n$ and entropy per particle $s$. Formally, the two approaches are related by a change of the canonical variables in the context of Lagrangian or Hamiltonian mechanics, as explained in~\cite{Brown_1993}.

Before continuing, it is worth briefly commenting on the relation to the pullback formalism, and the closely related effective field theory (EFT) approach, which provide geometrical treatments of relativistic perfect fluids. In the pullback formulation, fluid elements are labelled by comoving scalar fields $\alpha^A$ or $\phi^A$, and the relevant conserved current is constructed as the dual of the pullback of a fluid-space volume form~\cite{Carter:1987qr,Andersson:2006nr}. Its conservation is therefore a geometric identity which holds off shell, rather than an on-shell consequence of the equations of motion. Closely related ideas appear in the modern EFT description of fluids, where the internal scalar field degrees of freedom $\phi^A$ are used to construct actions invariant under internal volume-preserving diffeomorphisms~\cite{Dubovsky:2011sj,Ballesteros:2012kv}. The conserved currents are then either a direct consequence of the pullback structure or arise as Noether currents associated with the internal symmetries~\cite{Dubovsky:2011sj,Ballesteros:2016kdx}. These approaches are therefore more closely related to the Brown rather than the Schutz formalism; in particular, they can be viewed as extensions of the reduced Lagrangian-coordinate description discussed in Section~5 of Brown's original work~\cite{Brown_1993}, which effectively integrates out the Lagrange multiplier constraints. For minimally coupled fluids, these formalisms provide equivalent geometric ways of describing the same dynamics. In non-minimal $f(R,L_{\rm m})$ models, however, the crucial issue remains as to which off-shell thermodynamic potential is chosen as $L_{\rm m}$. For the remainder of this paper, we focus only on the Schutz and Brown actions, while noting that these alternative formulations are also widely used in the literature.

For each approach, we will introduce the relevant variables and construct the proper fluid action, deriving the field equations and fluid's equations of motion. We will also study the modified conservation equations and discuss the effective dark sector interactions. These will be presented in Section~\ref{Schutz_approach} for Schutz's approach and Section~\ref{Browns_approach} for Brown's approach. Further discussions and implications are presented in Section~\ref{Implications}.

\paragraph{Notation and conventions.}
We follow the convention of~\cite{Misner:1973prb} by setting the metric signature to $\{-,+,+,+\}$. We set $c=1$ and define $\kappa=8\pi G$ where $G$ is Newton's gravitational constant, and assume Einstein's summation notation throughout the paper.

\section{Schutz's approach}
\label{Schutz_approach}

A comprehensive relativistic perfect fluid Lagrangian built from velocity potentials was presented by Schutz~\cite{Schutz:1970} in 1970. The Lagrangian density is $\mathcal{L}_{\rm m} = p(\upmu,s)\sqrt{-g}$, with the pressure being the primary thermodynamic quantity. The entropy per particle is an independent variable, but the chemical potential is not. It is defined in terms of the other independent velocity potentials $\phi, \alpha, \beta,$ and $\theta$ as
\begin{align}
    \upmu^2 = -V^\nu V_\nu\,, \quad \textrm{where} \quad V_\nu =  \phi_{,\nu}+\alpha \beta_{,\nu}+\theta s_{,\nu} \,.
\end{align}
The term $V^\nu$ is often called the Taub current, which is defined as $V_\nu = \upmu U_\nu$. Here $U_\nu$ is the fluid's 4-velocity as in~\eqref{Fluid_T}. The action in this approach is given by
\begin{align}
\label{Schutz_action}
   S = \frac{1}{2\kappa}\int f(R,p)\sqrt{-g}\, d^4x+\int p\sqrt{-g}\,d^4x
\end{align}
Comparing the action with the general action~\eqref{general_action}, we have $\mathcal{C}_{\rm m} = 0$. The independent velocity potentials are implicit in the pressure and chemical potential $p(\upmu,s)$ and the equations of motion are obtained by varying these expressions with respect to the independent velocity potentials. The independent variables in this setup are 
\begin{align}
    \{ g^{\mu\nu}, \alpha, \beta, \phi, \theta, s \}\,.
\end{align}
We count a total of 15 independent variables, 10 encoded in the metric tensor and 5 fluid variables.

\subsection{Variations and gravitational field equations}

Varying the action~\eqref{Schutz_action} gives
\begin{align}
    \delta S = \frac{1}{2\kappa}\int \Big(f_R \delta R +f_L \delta p -\frac{1}{2}fg_{\mu\nu}\delta g^{\mu\nu}\Big)\sqrt{-g}d^4x+\int \Big( \delta p-\frac{1}{2}pg_{\mu\nu}\delta g^{\mu\nu}\Big)\sqrt{-g}d^4x \,,
\end{align}
where $f_R$ and $f_L$ denotes the derivatives of $f$ with respect to $R$ and $L_m$, respectively. The Ricci scalar is a purely geometric quantity that depends only on the metric, and its variation $\delta R$ gives the well-known expression
\begin{align}
    \delta R = \big(R_{\mu\nu}+g_{\mu\nu}\Box-\nabla_\mu \nabla_\nu \big)   \delta g^{\mu\nu} \,.
\end{align}

The variation of the pressure has to be performed carefully, in order to take into account all of the velocity potentials. This was calculated in~\cite{Boehmer:2025afy}, so it suffices to state the result here as
\begin{align}
    \delta p = -\frac{1}{2}\upmu \frac{\partial p}{\partial \upmu}U_\mu U_\nu\delta g^{\mu\nu}-\frac{\partial p}{\partial \upmu}U^\nu\Big(\delta \phi_{,\nu}+\beta_{,\nu}\delta \alpha+\alpha \delta \beta_{,\nu}+s_{,\nu}\delta \theta+\theta\delta s_{,\nu} \Big) + \frac{\partial p}{\partial s}\delta s \,.
\end{align}
Thus, the total action variation can be explicitly written as
\begin{multline}
\label{Schutz_full_action} 
    \delta S = \frac{1}{2\kappa}\int \bigg[ \Big(\big(R_{\mu\nu} + g_{\mu\nu}\square - \nabla_\mu \nabla_\nu \big)f_R-\frac{1}{2}fg_{\mu\nu} \Big)\delta g^{\mu\nu}
    -\frac{1}{2}f_L\upmu\frac{\partial p}{\partial \upmu} U_\mu U_\nu\delta g^{\mu\nu}\\-f_L\frac{\partial p}{\partial \upmu}U^\nu\Big(\delta \phi_{,\nu}+\beta_{,\nu}\delta \alpha+\alpha \delta \beta_{,\nu}+s_{,\nu}\delta \theta+\theta\delta s_{,\nu} \Big)+f_L\frac{\partial p}{\partial s}\delta s\bigg] \sqrt{-g}\,d^4x\\
    -\int \bigg[\frac{1}{2}\upmu\frac{\partial p}{\partial \upmu} U_\mu U_\nu\delta g^{\mu\nu}+\frac{\partial p}{\partial \upmu}U^\nu\Big(\delta \phi_{,\nu}+\beta_{,\nu}\delta \alpha+\alpha \delta \beta_{,\nu}+s_{,\nu}\delta \theta+\theta\delta s_{,\nu} \Big)\\ 
    - \frac{\partial p}{\partial s}\delta s+\frac{1}{2}pg_{\mu\nu}\delta g^{\mu\nu} \bigg]\sqrt{-g}\,d^4x \,.
\end{multline}
By collecting the terms of the metric variation $\delta g^{\mu\nu}$, we get the gravitational field equations
\begin{align}
\label{Schutz_field_eqns}
    f_RR_{\mu\nu} + \big(g_{\mu\nu}\square - \nabla_\mu \nabla_\nu \big)f_R-\frac{1}{2}fg_{\mu\nu}=\kappa T_{\mu\nu}+\frac{1}{2}f_L\upmu\frac{\partial p}{\partial \upmu} U_\mu U_\nu \,,
\end{align}
where $T_{\mu\nu}$ is the fluid energy-momentum tensor given by~\eqref{Fluid_T}. We can further write the right-hand side of~\eqref{Schutz_field_eqns} in terms of an \textit{effective} energy-momentum tensor as
\begin{align}
\label{schutz_effective_field_eqs}
     f_RR_{\mu\nu} + \big(g_{\mu\nu}\square - \nabla_\mu \nabla_\nu \big)f_R-\frac{1}{2}fg_{\mu\nu}=\kappa T_{\mu\nu}^{\rm eff} \,,
\end{align}
where
\begin{align}
\label{effective_T_schutz}
    T_{\mu\nu}^{\rm eff} := T_{\mu\nu}+\frac{1}{2\kappa}f_L\upmu\frac{\partial p}{\partial \upmu} U_\mu U_\nu \,.
\end{align}
Using~\eqref{Fluid_T}, the effective energy-momentum tensor defines the effective energy density given by
\begin{align}
\label{effective_density_schutz}
   T_{\mu\nu}^{\rm eff} := \rho_{\rm eff} U_\mu U_\nu + ph_{\mu\nu} \,, \quad \textrm{where} \quad  \rho_{\rm eff} := \rho +\frac{1}{2\kappa}f_L\upmu\frac{\partial p}{\partial \upmu}\,.
\end{align}

There are two points to note in this result. First, in~\eqref{schutz_effective_field_eqs} one may be tempted to conclude that the geometric terms reside entirely on the left-hand side of the equation while the matter terms sit on the right-hand side. However, this clearly is not the case for arbitrary non-minimal curvature-matter couplings, where all terms depending on $f$ and its derivative can in principle contain both matter and curvature sources. In the special case of minimal curvature-matter couplings $f_{RL} \equiv\partial^2 f/\partial R\partial L_m =0$, we retrieve $f(R)$ gravity with an effective thermodynamic quantity, namely the energy density $\rho_{\rm eff}$. 

Second, we note that the additional effective term in~\eqref{effective_T_schutz} is along the fluid flow direction, which allowed for the introduction of the effective energy density term. This is a direct result from~\eqref{Fluid_T}, for relativistic perfect fluids, where the energy flows along the fluid flow direction while the pressure acts in the direction perpendicular to the fluid's flow. This implies the simple result
\begin{align}
    h^{\sigma\mu}T_{\mu\nu}^{\rm eff} = h^{\sigma\mu} T_{\mu\nu} \,.
\end{align}

\subsection{Fluid equations and thermodynamic quantities}
\label{Fluid equations Schutz}

The fluid's equations of motion are obtained from all the variations in~\eqref{Schutz_full_action} besides $\delta g^{\mu\nu}$. Collecting the relevant terms for the variation of each independent variable gives
\begin{alignat}{2}
    \delta\alpha:& \qquad & \bigl(1+\frac{f_L}{2\kappa}\bigr)\frac{\partial p}{\partial \upmu}U^\nu\beta_{,\nu}  &= 0 \,,  \\
    \delta\beta:& \qquad &  \partial_\nu\Bigl(\bigl(1+\frac{f_L}{2\kappa}\bigr)\frac{\partial p}{\partial \upmu}U^\nu\alpha\sqrt{-g}\Bigr) &= 0 \,,  \\
    \delta\phi:& \qquad & \partial_\nu\Bigl(\bigl(1+\frac{f_L}{2\kappa}\bigr)\frac{\partial p}{\partial \upmu}U^\nu\sqrt{-g}\Bigr) &= 0 \,, \label{delta_phi_schutz_relation}\\
    \delta\theta:& \qquad &\bigl(1+\frac{f_L}{2\kappa}\bigr)\frac{\partial p}{\partial \upmu}U^\nu s_{,\nu} &= 0 \,,  \\
    \delta s:& \qquad & \bigl(1+\frac{f_L}{2\kappa}\bigr)\frac{\partial p}{\partial s}\sqrt{-g}+ \partial_\nu\Bigl(\bigl(1+\frac{f_L}{2\kappa}\bigr)\frac{\partial p}{\partial \upmu}U^\nu\theta\sqrt{-g}\Bigr) \label{delta_s_schutz1} &= 0 \,. 
\end{alignat}
Note that when $1+\frac{f_L}{2\kappa} = 0$, the above equations are identically true. 
In this degenerate case, we can directly solve the condition $1+\frac{f_L}{2\kappa} = 0$ to give $f(R,L_{\rm m})= -2\kappa L_{\rm m} + f^{(1)}(R)$, where $f^{(1)}(R)$ is an arbitrary function of the Ricci scalar. This is clearly just pure $ f^{(1)}(R)$ gravity with no matter source, due to the cancelling of the two $L_{\rm m}$ terms in the total action. Consequently we will assume $1+\frac{f_L}{2\kappa} \neq 0$ throughout this work to avoid this trivial case.

We notice that the fluid's equations of motion also pick up an additional term that depends on the coupling function $f(R,L_{\rm m})$. However, this additional term can be added to the definition of the particle number density (see Eq.~\eqref{thermodynamic_Schutz}) to define an effective particle number density term:
\begin{align}
\label{n_schutz}
    n_{\rm eff} =  
    \Big(1 + \frac{f_L}{2\kappa}\Big)\frac{\partial p}{\partial \upmu} = 
    \Big(1 + \frac{f_L}{2\kappa}\Big)n \,,
    \qquad n = \frac{\partial p}{\partial \upmu} \,.
\end{align}
The fluid's equations now take the convenient form given by
\begin{alignat}{2}
    \delta\alpha:& \qquad & U^\nu \beta_{,\nu}  &= 0 \,, \label{delta_alpha_schutz} \\
    \delta\beta:& \qquad &  U^\nu \alpha_{,\nu} &= 0 \,, \label{delta_beta_schutz} \\
    \delta\phi:& \qquad & \partial_\nu(\sqrt{-g}\,n_{\rm eff}U^\nu) &= 0 \,, \label{delta_phi_schutz}\\
    \delta\theta:& \qquad &U^\nu s_{,\nu}  &= 0 \,, \label{delta_theta_schutz} \\
    \delta s:& \qquad & nU^\nu\theta_{,\nu} + \frac{\partial p}{\partial s} &= 0 \,. \label{delta_s_schutz}
\end{alignat} 
We have a total of five fluid equations of motion as expected. We see that Eqs.~\eqref{delta_alpha_schutz}--\eqref{delta_beta_schutz}, and~\eqref{delta_theta_schutz}--\eqref{delta_s_schutz} are independent of the non-minimal curvature-matter coupling. Hence, these equations retain their general relativistic form. Thus, the number conservation equation~\eqref{delta_phi_schutz} is the only fluid's equation of motion affected by the non-minimal coupling with matter. One could then naturally introduce an effective vector-density particle-number flux $J_{\rm eff}^\mu = \sqrt{-g}\,n_{\rm eff}U^\mu$ which satisfies the usual conservation equation $\partial_\mu J_{\rm eff}^\mu = 0$.

On the other hand, if we use the standard number density term then from equation ~\eqref{delta_phi_schutz_relation} we obtain the interesting relation
\begin{equation}
\label{schutz_continuity_eqn}
    \nabla_\nu (nU^\nu) = - nU^\nu\nabla_\nu \log\Bigl(1+\frac{f_L}{2\kappa}\Bigr) \,.
\end{equation}
The result indicates that particle number is not necessarily conserved in the direction of the flow and that it depends on the curvature-matter coupling $f_L$. Only the effective number density $n_{\rm eff}$ is conserved, which can also be written as $\nabla_\nu(n_{\rm eff} U^\nu)=0$.

\subsection{On-shell argument}
\label{On-shell Argument Schutz}

One often states that, on shell, the Lagrangian can be defined as either the pressure or the negative energy density. For minimally coupled fluids one has indeed $L_{\rm m}=p=-\rho$, on shell (up to boundary terms). But it has been noted, as stated before, that this is not the case in models with non-minimal curvature-matter couplings as they lead to different physical results. Let us consider the fluid action
\begin{equation}
    S_{\rm fluid}=\int L_{\rm m} \sqrt{-g}\, d^4x  = \int p\, \sqrt{-g}\,d^4x \,.
\end{equation}
Using our definitions and the fluid equations, the action can be written using~\eqref{density_relation_Schutz} as 
\begin{align}
\label{onshell_action1}
    S_{\textrm{fluid}} &=  \int \Bigl(\upmu\frac{\partial p}{\partial \upmu} -\rho \Bigr) \sqrt{-g}\,d^4x  \notag \\
    &= \int \Bigl(\upmu\frac{\partial p}{\partial \upmu} -\rho + \frac{f_L}{2\kappa}\upmu\frac{\partial p}{\partial \upmu} - \frac{f_L}{2\kappa}\upmu\frac{\partial p}{\partial \upmu} \Bigr) \sqrt{-g}\,d^4x  \notag\\
    &= \int \Bigl(\upmu \bigl(\frac{f_L}{2\kappa}+1\bigr)\frac{\partial p}{\partial \upmu} - \bigl(\rho  + \frac{f_L}{2\kappa}\upmu\frac{\partial p}{\partial \upmu} \bigr) \Bigr) \sqrt{-g}\,d^4x  \notag\\
    &=\int \upmu n_{\rm eff} \sqrt{-g}\,d^4x-\int\rho_{\textrm{eff}}\sqrt{-g}\,d^4x \,,
\end{align}
where we have used the definitions~\eqref{effective_density_schutz} and~\eqref{n_schutz}. We have that $V_\nu = \upmu U_\nu$ and thus we can also write $\upmu = -U^\nu(\phi_{,\nu}+\alpha\beta_{,\nu}+\theta s_{,\nu})$. Using~\eqref{delta_alpha_schutz} and~\eqref{delta_theta_schutz}, this reduces to $\upmu = -U^\nu\phi_{,\nu}$. Substituting this back into the first term of~\eqref{onshell_action1} gives
\begin{align}
     \int \upmu n_{\rm eff} \sqrt{-g}\,d^4x &= -\int U^\nu\phi_{,\nu} n_{\rm eff} \sqrt{-g}\,d^4x \notag \\
    &= - \int \partial_{\nu} \bigl(\sqrt{-g}  U^\nu\phi n_{\rm eff} \bigr) \,d^4x + \int \phi \partial_{\nu} \bigl(\sqrt{-g} U^\nu n_{\rm eff}  \bigr)\,d^4x  = 0 \,,
\end{align}
because the first term is a total derivative and the second term vanishes by~\eqref{delta_phi_schutz}. Consequently, we end up with the on-shell Lagrangian
\begin{align}
    S_{\rm fluid} = \int p\, \sqrt{-g}\,d^4x = -\int\rho_{\rm eff}\sqrt{-g}\,d^4x 
    \qquad\text{on shell} \,.
\end{align}
One immediately sees that on shell, the equivalence is established between the pressure and the \textit{effective} energy density. This equivalence therefore holds for the fluid action $L_m$, but not inside the non-minimally coupled function itself, since
\begin{align}
\label{onShell_nonequiv}
    \int f(R,p)\sqrt{-g}\,d^4x \neq \int f(R,-\rho_{\rm eff}) \sqrt{-g}\,d^4x 
    \qquad\text{on shell} \,.
\end{align}
This result follows from the fact that integration by parts was needed (to eliminate boundary terms) in the derivation.

\subsection{Conservation equations}
\label{sec_Schutz_conservtion_eqs}

As stated earlier, a special property of non-minimally curvature-matter coupled models is that they lead to non-geodesic motion of free particles due to the non-conservation of energy-momentum tensor. There are two ways to derive the conservation equations: one can take the covariant derivative of the field equations and derive the result for $\nabla^\mu T_{\mu\nu}$, or one can start from the fluid equations~\eqref{Fluid_T} and take their covariant derivative. To show the consistency of the results, we shall derive the equations using both approaches.

\subsubsection{Conservation equations from the field equations}

If we take the covariant derivative of the field equations in~\eqref{Schutz_field_eqns}, we get
\begin{multline}
\label{Schutz_cons1}
    f_R\nabla^\mu R_{\mu\nu} + R_{\mu\nu} \nabla^\mu f_{R}+ \big(\nabla_\nu\square - \square \nabla_\nu \big)f_R-\frac{1}{2} f_R \nabla_\nu R\\-\frac{1}{2} f_L \nabla_\nu L_{\rm m}=\kappa \nabla^\mu \Bigl( T_{\mu\nu}+\frac{f_L}{2\kappa}\upmu\frac{\partial p}{\partial \upmu} U_\mu U_\nu \Bigr) \,.
\end{multline}
The first four terms on the left-hand side vanish geometrically, and identifying the effective energy-momentum tensor gives
\begin{align}
\label{effective_cons_eqn}
    \nabla^\mu T^{\rm eff}_{\mu\nu} &= -\frac{f_L}{2\kappa} \nabla_\nu L_{\rm m}\,.
\end{align}
Notice that this equation is general for any $f(R,L_{\rm m})$ model with an action~\eqref{general_action} in which we can identify a $T^{\rm eff}_{\mu\nu}$ term. In Schutz's approach, $L_{\rm m}=p(\upmu,s)$. Writing out $T^{\rm eff}_{\mu\nu}$ explicitly gives
\begin{align}
    \nabla^\mu T_{\mu\nu} &= -\frac{f_L}{2\kappa} \frac{\partial p}{\partial s}\nabla_\nu s -\frac{f_L}{2\kappa}g_{\mu\nu} \frac{\partial p}{\partial \upmu}\nabla^\mu \upmu  - \nabla^\mu \Bigl(\frac{f_L}{2\kappa}\upmu\frac{\partial p}{\partial \upmu} U_\mu U_\nu \Bigr) \notag\\
     &= -\frac{1}{2\kappa}\Bigl(f_L \frac{\partial p}{\partial s}\nabla_\nu s + h_{\mu\nu}f_L \frac{\partial p}{\partial \upmu} \nabla^\mu \upmu + \upmu\nabla^\mu \bigl(f_L \frac{\partial p}{\partial \upmu}U_\mu U_\nu \bigr) \Bigr)\,.
\end{align}
Contracting along the fluid flow leads to
\begin{align}
    U^\nu\nabla^\mu T_{\mu\nu} &= -\frac{1}{2\kappa}\Bigl(f_L \frac{\partial p}{\partial s}U^\nu\nabla_\nu s + h_{\mu\nu}f_L \frac{\partial p}{\partial \upmu}U^\nu \nabla^\mu \upmu + \upmu U^\nu \nabla^\mu \bigl(f_L \frac{\partial p}{\partial \upmu}U_\mu U_\nu \bigr) \Bigr) \notag \\
   &= \frac{1}{2\kappa} \upmu \nabla^\mu \bigl(f_L \frac{\partial p}{\partial \upmu}U_\mu\bigr) \neq 0
    \,,
    \label{Schutz_parallel_contract_1}
\end{align}
where we have invoked~\eqref{delta_theta_schutz}, $U^\nu h_{\mu\nu}=0$, $U_\nu U^\nu = -1$, and $U^\nu \nabla^\mu U_\nu = 0$. Furthermore, relation~\eqref{delta_phi_schutz} allows us to write
\begin{align}
\label{Schutz_parallel_contract_2}
    U^\nu\nabla^\mu T_{\mu\nu} &= - \upmu \nabla^\mu \bigl( \frac{\partial p}{\partial \upmu}U_\mu \bigr) \neq 0 \,.
\end{align}
This final equation is nothing but the usual number flux conservation equation, which we showed to be non-vanishing in the presence of matter-curvature couplings~\eqref{schutz_continuity_eqn}. This demonstrates that the non-conservation of the energy-momentum tensor in the Schutz formalism is a direct consequence of particle-number conservation violation.

On the other hand, contraction in the direction perpendicular to the flow gives
\begin{align}
    h^\nu_\sigma\nabla^\mu T_{\mu\nu} &= -\frac{1}{2\kappa}\Bigl(h_{\mu\sigma}f_L \frac{\partial p}{\partial s}\nabla^\mu s + h_{\mu\sigma}f_L \frac{\partial p}{\partial \upmu} \nabla^\mu \upmu + \upmu h^\nu_\sigma \nabla^\mu \bigl(f_L \frac{\partial p}{\partial \upmu}U_\mu U_\nu \bigr) \Bigr) \notag \\
    &= -\frac{1}{2\kappa}\Bigl(f_L \frac{\partial p}{\partial s}\nabla_\sigma s + h_{\mu\sigma}f_L \frac{\partial p}{\partial \upmu} \nabla^\mu \upmu + \upmu f_L \frac{\partial p}{\partial \upmu}U_\mu h^\nu_\sigma \nabla^\mu U_\nu \Bigr) \,.
    \,
\end{align}
Now using~\eqref{delta_s_schutz} and~\eqref{n_schutz}, this reduces to
\begin{align}
    h^\nu_\sigma\nabla^\mu T_{\mu\nu} &= -\frac{1}{2\kappa}\Bigl(-f_L \frac{\partial p}{\partial \upmu}U^\lambda \nabla_\lambda \theta\nabla_\sigma s + h_{\mu\sigma}f_L \frac{\partial p}{\partial \upmu} \nabla^\mu \upmu + \upmu f_L \frac{\partial p}{\partial \upmu}U_\mu h^\nu_\sigma \nabla^\mu U_\nu \Bigr) \notag \\
     &= \frac{f_L}{2\kappa}\frac{\partial p}{\partial \upmu}h^\nu_\sigma\Bigl( U^\lambda \nabla_\lambda \theta\nabla_\nu s - h_{\mu\nu}  \nabla^\mu \upmu - \upmu U_\mu \nabla^\mu U_\nu \Bigr)
    \,.
\end{align}
The term in the parentheses vanishes when contracted with $h^\nu_\sigma$ (see again~\cite{Boehmer:2025afy}) and thus
\begin{align}
\label{Schutz_perpen_contraction_1}
     h^\nu_\sigma\nabla^\mu T_{\mu\nu} &= 0 \,.
\end{align}
We find that the energy-momentum tensor is conserved in the perpendicular direction but not along the flow direction. A discussion of this result and its relation to the result in~\eqref{n_schutz} is presented in Section~\ref{Implications}. 
We now derive the conservation equations from the fluid equation.

\subsubsection{Conservation equations from the fluid equation}

Starting by taking the covariant derivative of~\eqref{Fluid_T} using Schutz's approach, we have
\begin{align}
\label{fluid_div}
    \nabla^\mu T_{\mu\nu} = \nabla^\mu \Big(\upmu\frac{\partial p}{\partial \upmu}U_\mu U_\nu\Big)+\nabla_\nu p \,.
\end{align}
The contraction along the fluid flow is trivial and yields
\begin{align}
    U^\nu \nabla^\mu T_{\mu\nu} &= -\nabla^\mu \Big(\upmu \frac{\partial p}{\partial \upmu} U_\mu \Big) + \frac{\partial p}{\partial \upmu}U^\nu \nabla_\nu \upmu\, + \frac{\partial p}{\partial s} U^\nu \nabla_\nu s \notag \\
    &= -\upmu \nabla^\mu \Big(\frac{\partial p}{\partial \upmu}U_\mu \Big) \neq 0\,,
\end{align}
which matches the result in~\eqref{Schutz_parallel_contract_2}. Note that this is not zero because of Eq.~\eqref{delta_phi_schutz}.

For the perpendicular contraction, we contract~\eqref{fluid_div} with $h^\nu_\sigma$: 
\begin{align}
    h^\nu_\sigma\nabla^\mu T_{\mu\nu} &= h^\nu_\sigma \nabla^\mu \Big(\upmu\frac{\partial p}{\partial \upmu}U_\mu U_\nu\Big)+\frac{\partial p}{\partial \upmu}h^\nu_\sigma\nabla_\nu \upmu +  \frac{\partial p}{\partial s}h^\nu_\sigma\nabla_\nu s \notag \\
    &= \upmu\frac{\partial p}{\partial \upmu}h^\nu_\sigma U_\mu \nabla^\mu U_\nu+\frac{\partial p}{\partial \upmu}h^\nu_\sigma \nabla_\nu \upmu +  \frac{\partial p}{\partial s}h^\nu_\sigma\nabla_\nu s \label{special_identity_1}\,.
\end{align}
We can now use the equation of motion~\eqref{delta_s_schutz} and the definition of $n$ to substitute for $\partial p/\partial s$ and write
\begin{align}
    h^\nu_\sigma\nabla^\mu T_{\mu\nu} &= \frac{\partial p}{\partial\upmu}h^\nu_\sigma \Big(\upmu U_\mu \nabla^\mu U_\nu + \nabla_\nu \upmu - U^\lambda\theta_{,\lambda}\nabla_\nu s \Big) =0\,,
\end{align}
as we found earlier in~\eqref{Schutz_perpen_contraction_1}.
We have therefore demonstrated consistent results in deriving the conservation equations for a relativistic perfect fluid using Schutz's approach. We find that in the case of non-minimal coupling between curvature and matter, the energy-momentum tensor is conserved in the perpendicular direction but not along the fluid flow. This is a consistent result with~\eqref{schutz_continuity_eqn}, where the number conservation equation is modified by the non-minimal coupling term.

\subsection{Effective equations and interactions} 
\label{Schutz_dark_intr}

Models with non-minimal couplings between curvature and matter can describe effective interactions in the dark sector, which have been motivated by recent cosmological observations~\cite{Li:2026xaz}. While one would often investigate whether a specific model can describe viable dark sector interactions, one should also investigate the effect of the relativistic fluid approach used.
In the present framework, the fundamental description is of a single relativistic fluid non-minimally coupled to curvature, which we can then identify as distinct effective fluid components. It should be noted, however, that this split is arbitrary and a matter of interpretation.

A simple dark sector interaction model can be studied by setting $f(R,L_m) = R + \bar{f}(R,L_m)$ so that $f_R=1+\bar{f}_R$ and $f_L = \bar{f}_L$, which isolates the Einstein tensor. This gives the field equations
\begin{align}
\label{eff_Schutz1}
    G_{\mu\nu} &=\kappa T_{\mu\nu}+\frac{1}{2}\bar{f}_L\upmu\frac{\partial p}{\partial \upmu} U_\mu U_\nu - \bigl(R_{\mu\nu} + g_{\mu\nu}\square - \nabla_\mu \nabla_\nu \bigr)\bar{f}_R + \frac{1}{2}\bar{f}g_{\mu\nu} \,.
\end{align}
By resorting to the definition of the effective energy-momentum term, we can write
\begin{align}
    G_{\mu\nu} &= \kappa \bigl(T^{\rm eff}_{\mu\nu}+\hat{T}_{\mu\nu}^{\bar{f}} \bigr) \qquad \textrm{where} \qquad \hat{T}_{\mu\nu}^{\bar{f}} := \frac{1}{2\kappa}\bar{f}g_{\mu\nu} - \frac{1}{\kappa} \bigl(R_{\mu\nu} + g_{\mu\nu}\square - \nabla_\mu \nabla_\nu \bigr)\bar{f}_R  \,.
\end{align}
The term $\hat{T}^{\bar{f}}_{\mu\nu}$ can be associated with geometrically-sourced dark energy. By associating the $T^{\rm eff}_{\mu\nu}$ term with a dark matter fluid sector, which maintains a geometric dependence via $\bar{f}_L$, we can identify the dark sector interaction by
\begin{align}
    \nabla^\mu T^{\rm eff}_{\mu\nu} &= - \nabla^\mu \hat{T}_{\mu\nu}^{\bar{f}}= 
    - \frac{1}{2\kappa}\bar{f}_{ L}\nabla_\nu L_{\rm m} =: \mathcal{Q}_\nu \,.
\end{align}
In Schutz's fluid approach, the interaction is given by a pressure gradient
\begin{align}
    \mathcal{Q}_\nu &= - \frac{1}{2\kappa}\bar{f}_{L}\nabla_\nu p \,.
    %= - \frac{1}{2\kappa}\bar{f}_{ L} \bigl(\frac{\partial p}{\partial \upmu}\nabla_\nu \upmu +\frac{\partial p}{\partial s}\nabla_\nu s \bigr) \,.
\end{align}
Physically, a term of this form is well-motivated. Pressure gradients are the force terms in the Euler equations, so it is unsurprising that it is the pressure gradient which drives the energy exchange between the fluid components.  

\section{Brown's approach}
\label{Browns_approach}

We now follow the same procedure to derive the field equations, fluid's equations of motion, conservation equations and dark sector interactions using Brown's fluid Lagrangian with Lagrange multipliers~\cite{Brown_1993}. Non-minimal curvature-matter couplings of $f(R,L_{\rm m})$ have been studied before using Brown's approach (see~\cite{Bertolami_2008, Bertolami:2008zh}). In the following, we derive the complete set of equations using the off-shell Lagrangian, allowing us to compare the results with Schutz's approach using velocity potentials. To begin with, we set 
\begin{align}
   \mathcal{L}_{\rm m}=-\rho(n,s)\sqrt{-g}\,, \quad \textrm{and} \quad \mathcal{C}_{\rm m} = J^\mu(\varphi_{,\mu}+s\theta_{,\mu}+\beta_A\alpha^A_{,\mu}) \,.
\end{align}
Here
\begin{itemize}
    \item  $J^\mu$ is the vector-density particle-number flux, which is related to $n$ by
    \begin{equation}
        J^\mu=\sqrt{-g}nU^\mu\,, \qquad 
        |J|=\sqrt{-g_{\mu\nu}J^\mu J^\nu}\,, \qquad 
        n=|J|/\sqrt{-g} \,, 
        \label{defnofJandn}
    \end{equation}
    and $U^\mu$ is the fluid's four-velocity satisfying $U_\mu U^\mu=-1$;
    \item  $\varphi$, $\theta$, and $\beta_A$ are all Lagrange multipliers with $A$ taking the values 1, 2, and 3, and $\alpha^A$ are the Lagrangian coordinates of the fluid; and
    \item the energy density $\rho(n,s)$ is the fundamental quantity and the pressure $p(n,s)$ is the derived quantity defined by 
    \begin{align}
        p = n\frac{\partial \rho}{\partial n} - \rho \,.
        \label{pressuredef}
    \end{align}
\end{itemize}

The action is given by
\begin{align}
    S =  \frac{1}{2\kappa}  \int  f(R,-\rho)\sqrt{-g}  d^4x + \int   \Bigl[ - \rho(n,s)\sqrt{-g} +J^\mu(\varphi_{,\mu}+s\theta_{,\mu}+\beta_A\alpha^A_{,\mu})\Bigr] d^4x \, .
\end{align}
Note that the number density $n$ is not an independent variable, but depends on the metric $g^{\mu\nu}$ and the particle number flux density $J^\mu$ as introduced in~\eqref{defnofJandn}. The independent variables, which we vary the action with respect to, are
\begin{align}
    \{ g^{\mu\nu}, J^\mu, \alpha^A, \beta_A, \varphi, \theta, s\} \,.
\end{align}
We count a total of 23 independent components: 10 encoded in the metric tensor, and 13 fluid variables. This much larger number, compared to the 15 in Schutz, requires additional equations of motion. This observation highlights that both approaches follow very different pathways in their setup of the variational approach to the fluid Lagrangian.

\subsection{Variations and gravitational field equations}

One advantage of Brown's approach is that all the independent variables appear explicitly in the Lagrangian. Variations with respect to all the independent variables yield
\begin{multline}
\label{Brown_action_var1}
     \delta S = \frac{1}{2\kappa}\int \Big(
    f_R \delta R - f_L \delta \rho -\frac{1}{2}fg_{\mu\nu}\delta g^{\mu\nu} \Big)\sqrt{-g}\, d^4x -\int \Big(\delta\rho-\frac{1}{2}\rho g_{\mu\nu}\delta g^{\mu\nu} \Big)\sqrt{-g}\,d^4x \\
    +\int\Big( (\varphi_{,\mu}+s\theta_{,\mu}+\beta_A\alpha^A_{,\mu})\delta J^\mu+J^\mu(\delta \varphi_{,\mu}+\theta_{,\mu}\delta s+ s\delta \theta_{,\mu}+\alpha^A_{,\mu}\delta\beta_A+\beta_A\delta\alpha^A_{,\mu})\Big)d^4x \,.
\end{multline}
The variations in the second line are already written in terms of independent variables and their derivatives. The only new variation here is $\delta \rho(n,s)$, which is given by
\begin{align}
    \delta \rho=\frac{\partial \rho}{\partial n}\delta n+\frac{\partial\rho}{\partial s}\delta s = \frac{1}{2}n\frac{\partial \rho}{\partial n}\Bigl(g_{\mu\nu} +
    U_\mu U_\nu \Bigr) \delta g^{\mu\nu}-
    \frac{\partial \rho}{\partial n} \frac{U_\mu}{\sqrt{-g}}\delta J^\mu +\frac{\partial\rho}{\partial s}\delta s \,.
\end{align}
Putting all these variations together and integrating by parts the relevant terms, yields
\begin{multline}
\label{Brown_action_var2}
    \delta S= \frac{1}{2\kappa} \int \bigg[\big(R_{\mu\nu} + g_{\mu\nu}\square - \nabla_\mu \nabla_\nu \big)f_R \delta g^{\mu\nu} \\
     - f_L n\frac{\partial \rho}{\partial n}\Big(\frac{1}{2} \big(g_{\mu \nu} + U_\mu U_\nu \big)\delta g^{\mu\nu} - \frac{U_\mu}{n\sqrt{-g}}\delta J^\mu \Big)  
      -f_L \frac{\partial \rho}{\partial s} \delta s - \frac{1}{2}fg_{\mu\nu}\delta g^{\mu\nu} \bigg]\sqrt{-g}\,d^4x
     \\- \int \bigg[\frac{1}{2}n\frac{\partial \rho}{\partial n}(g_{\mu\nu}+U_\mu U_\nu) \delta g^{\mu \nu}  - \frac{\partial \rho}{\partial n} \frac{U_{\mu}}{\sqrt{-g}} \delta J^{\mu} + \frac{\partial \rho}{\partial s} \delta s-\frac{1}{2}\rho g_{\mu\nu}\delta g^{\mu\nu}\bigg] \sqrt{-g}\,d^4x
     \\
    + \int\bigg[(\varphi_{,\mu}+s\theta_{,\mu}+\beta_A\alpha^A_{,\mu})\delta J^\mu - J^\mu_{\ ,\mu} \delta \varphi+J^\mu \theta_{,\mu}\delta s \\
    -\big(J^\mu s\big)_{,\mu}\delta \theta+ J^\mu \alpha^A_{,\mu}\delta \beta_A - \big(J^\mu \beta_A\big)_{,\mu} \delta \alpha^A \bigg] d^4x \,.
\end{multline}

Collecting the $\delta g^{\mu\nu}$ terms, and using Eq.~\eqref{Fluid_T} with the identification $\rho+p=n\partial \rho/\partial n$, gives us the metric field equations 
\begin{align}
\label{Brown_field_eqn}
     f_R R_{\mu\nu} + \big(g_{\mu\nu}\square - \nabla_\mu \nabla_\nu \big)f_R   - \frac{1}{2}fg_{\mu\nu} = \kappa T_{\mu\nu}+\frac{1}{2}f_L n\frac{\partial \rho}{\partial n} h_{\mu\nu} \,.
\end{align}
Analogously to Schutz's approach, the additional terms on the right-hand side act as modifications to the energy-momentum sources of the field equations. We now define an effective energy-momentum tensor by 
\begin{equation}
    \label{effective_T_Brown}
    T_{\mu \nu}^{\textrm{eff}} := T_{\mu\nu}+\frac{1}{2\kappa}f_L n\frac{\partial \rho}{\partial n} h_{\mu\nu} \,.
\end{equation}
Notice here, the additional term in $T_{\mu\nu}^{\rm eff}$ acts in the perpendicular direction of the flow and we have the result
\begin{equation}
    U^\mu T_{\mu \nu}^{\textrm{eff}} = U^\mu T_{\mu\nu} \,.
\end{equation}
Therefore we can associate this additional term with the fluid's pressure and define an effective pressure, similar to~\eqref{effective_density_schutz}. Here, we define
\begin{equation}
    p_{\rm eff} := p + \frac{1}{2\kappa}f_L n\frac{\partial \rho}{\partial n} \,. 
\end{equation}
It should be noted that while an analogy appears between Schutz's approach and Brown's approach in deriving the field equations and the emergence of effective quantities, they lead to distinct effects. Schutz's approach introduces an effective energy density term and Brown's approach introduces an effective pressure, each acting in a direction orthogonal to the other. The consequences of this modification will be further discussed in Sec.~\ref{Brown_conservation_equations} when the conservation equations are derived.

\subsection{Fluid equations and thermodynamic quantities}

The fluid's equations of motion from all the other variations in~\eqref{Brown_action_var2} are directly given by
\begin{alignat}{2}
    \delta J^\mu:& \qquad &\frac{1}{2\kappa}f_L \frac{\partial \rho}{\partial n}U_\mu+ \frac{\partial \rho}{\partial n}U_\mu +\varphi_{,\mu}+s\theta_{,\mu}+\beta_A\alpha^A_{,\mu} &= 0 \,,     \label{Jeq_brown}\\
    \delta\alpha^A:& \qquad &J^\mu \beta_{A,\mu} &= 0 \,, 
    \label{alphaeq_brown} \\
    \delta\beta_A:& \qquad &J^\mu \alpha^A_{, \mu} &= 0 \,, 
    \label{betaeq_brown} \\
    \delta\varphi:& \qquad & J^\mu_{,\mu} &= 0 \,,
    \label{phieq_brown} \\
    \delta\theta:& \qquad &J^\mu s_{,\mu} &= 0 \,, 
    \label{thetaeq_brown} \\
    \delta s:& \qquad & \frac{1}{2\kappa}f_L \frac{\partial \rho}{\partial s} + \frac{\partial \rho}{\partial s} - nU^\mu \theta_{,\mu}  &= 0 \,.
    \label{seq_brown}
\end{alignat}
In total, we have $13$ equations of motion, that is eight additional equations compared to the case of Schutz's approach. Equations~\eqref{alphaeq_brown}--\eqref{thetaeq_brown} are not affected by the non-minimal coupling and modifications are only observed in~\eqref{Jeq_brown} and~\eqref{seq_brown}. We also notice that both the number and entropy flux conservation equations~\eqref{phieq_brown}--\eqref{thetaeq_brown} take their standard form, implying the continuity equation for the fluid will be unchanged. Moreover, while the entropy flux is conserved in both Brown and Schutz's approaches, it is only the \textit{effective} number flux that is conserved in Schutz's approach.

Since the chemical potential is $\upmu=\partial\rho/\partial n$, we can define an effective chemical potential using~\eqref{Jeq_brown} by
\begin{align}
\label{effective_chemical_potential}
    \upmu_{\rm eff} := \Big(1+\frac{f_L}{2\kappa}\Big)\frac{\partial \rho}{\partial n} \,,
\end{align}
and rewrite the equation of motion as
\begin{align}
\label{Jeq_Brown_effective}
    -\upmu_{\rm eff}U_\mu=\varphi_{,\mu}+s\theta_{,\mu}+\beta_A\alpha^A_{,\mu} \,.
\end{align}
A thermodynamic relation from the entropy variation~\eqref{seq_brown} also emerges here, allowing us to define an effective temperature using $n\mathcal{T} = \partial \rho/\partial s$ (see Eq.~\eqref{thermodynamic_brown}), where 
\begin{align}
\label{temperature potential}
    U^\mu\theta_{,\mu} = \Bigl(1+\frac{f_L}{2\kappa}\Bigr)\frac{1}{n}\frac{\partial \rho}{\partial s} = \Bigl(1+\frac{f_L}{2\kappa}\Bigr)\mathcal{T} =: \mathcal{T}_{\rm eff} \,.
\end{align}
These results match the conclusions presented in~\cite{Bertolami_2008}.

If we further contract~\eqref{Jeq_brown} with $U^\mu$, we can then substitute the term $U^\mu \theta_{,\mu}$ into~\eqref{seq_brown}. Using~\eqref{betaeq_brown} and $U^\mu U_\mu=-1$, we get the relation 
\begin{align}
\label{effective_Helmholtz_brown}
    U^\mu \varphi_{,\mu} = \upmu_{\rm eff} - s\mathcal{T}_{\rm eff}:=\mathcal{F}_{\rm eff} \,,
\end{align}
where $\mathcal{F}_{\rm eff}$ is the effective Helmholtz free energy. In the absence of couplings, we retrieve the conventional thermodynamic relations in this approach presented in~\cite{Brown_1993}. These relations allow the identification of $\theta$ as a potential for temperature and $\varphi$ as the potential for the Helmholtz free energy. Evidently, the non-minimal coupling leads to an effective modification of these thermodynamic quantities. It should be noted, however, that other thermodynamic identifications are possible depending on convention. For instance, what we label as the chemical potential $\upmu$~\eqref{chemical_potential} is often called the enthalpy per particle. In that terminology, the combination $\upmu-s\mathcal{T}$ is interpreted as the chemical potential, and $\varphi$ is the corresponding chemical momentum; for further discussion and generalisations, see \cite{Iosifidis:2024ksa}.

\subsection{On-shell argument}
\label{On-shell Argument Brown}

We begin with the fluid Lagrangian and immediately see that
\begin{align}
    S_{\textrm{fluid}} &= \int (-\rho)\, \sqrt{-g}\,d^4x = 
    \int \Bigl(p-n\frac{\partial \rho}{\partial n}\Bigr) \sqrt{-g}\,d^4x  \notag \\
    &= \int \Bigl(p-n\frac{\partial \rho}{\partial n} + \frac{f_L}{2\kappa}n\frac{\partial \rho}{\partial n} - \frac{f_L}{2\kappa}n\frac{\partial \rho}{\partial n}\Bigr) \sqrt{-g}\,d^4x   \notag \\
    &=\int p_{\rm eff} \sqrt{-g}\, d^4x-\int n \upmu_{\rm eff} \sqrt{-g}\,d^4x \notag \\
    &= \int p_{\rm eff} \sqrt{-g}\, d^4x \,,
\end{align}
where we have also used 
\begin{equation}
    \int n \upmu_{\rm eff} \sqrt{-g}\,d^4x = 0 \,,
\end{equation}
which can be seen from the fluid's equations of motion. This means we have
\begin{align}
    \int (-\rho)\, \sqrt{-g}\,d^4x = \int p_{\rm eff} \sqrt{-g}\, d^4x 
    \qquad \text{on shell}\,.
\end{align}
It is also straightforward to show that, after integrating by parts and eliminating boundary terms, the constraint part vanishes $\int \mathcal{C}_m\, d^4x = 0$ on shell. However, when this matter Lagrangian is used inside the argument of an arbitrary function, one cannot make any statements about on-shell equivalence. In particular, one finds
\begin{align}
\label{onShell_nonequiv2}
    \int f(R,-\rho)\sqrt{-g}\,d^4x \neq \int f(R,p_{\rm eff}) \sqrt{-g}\,d^4x 
    \qquad\text{on shell} \,.
\end{align}
We thus stress that $-\rho$ and $p_{\rm eff}$ are not exchangeable, even on shell, for models with non-minimal curvature-matter coupling. 
\subsection{Conservation equations}
\label{Brown_conservation_equations}

The $h_{\mu\nu}$ component in the effective energy-momentum tensor~\eqref{effective_T_Brown} already indicates that, contrary to Schutz, Brown's approach produces a modification perpendicular to the direction of the fluid flow. It is thus natural to expect different conservation equations than that observed using Schutz's approach. We will derive the conservation equations in both directions and check for consistency by deriving the equations starting from the field equations~\eqref{Brown_field_eqn} and the fluid equation.

\subsubsection{Conservation equations from the field equations}
\label{conservation equations from the field equations}

One immediately finds that taking the covariant derivative of the field equations~\eqref{Brown_field_eqn} also leads to~\eqref{effective_cons_eqn}, this time with $T_{\mu\nu}^{\rm eff}$ given by~\eqref{effective_T_Brown} and $L_{\rm m}=-\rho$. We then have 
\begin{align}
    \nabla^\mu T_{\mu\nu} &= \frac{f_L}{2\kappa} g_{\mu\nu}\nabla^\mu \rho - \nabla^\mu \bigl(\frac{f_L}{2\kappa}n\frac{\partial \rho}{\partial n}h_{\mu\nu}\bigr) \notag \\
    &= \frac{f_L}{2\kappa} g_{\mu\nu}\frac{\partial \rho}{\partial s}\nabla^\mu s  + \frac{f_L}{2\kappa} g_{\mu\nu}\frac{\partial \rho}{\partial n}\nabla^\mu n  - \nabla^\mu \bigl(\frac{f_L}{2\kappa}n\frac{\partial \rho}{\partial n}h_{\mu\nu}\bigr) \notag \\
    &= \frac{f_L}{2\kappa} g_{\mu\nu} \frac{\partial \rho}{\partial s}\nabla^\mu s - \frac{f_L}{2\kappa} \frac{\partial \rho}{\partial n} U_\mu U_\nu\nabla^\mu n - n \nabla^\mu\bigl(\frac{f_L}{2\kappa}\frac{\partial \rho}{\partial n}h_{\mu\nu} \bigr) \,. \label{Brown_field_cons_3}
\end{align}
Taking the contraction along the fluid flow leads to 
\begin{align}
    U^\nu \nabla^\mu T_{\mu\nu} &=  \frac{f_L}{2\kappa}  \frac{\partial \rho}{\partial s}U^\nu\nabla_\nu s + \frac{f_L}{2\kappa} \frac{\partial \rho}{\partial n} U_\mu\nabla^\mu n + \frac{f_L}{2\kappa}n\frac{\partial \rho}{\partial n}\nabla^\mu U_\mu \,.
\end{align}
The first term vanishes from~\eqref{thetaeq_brown}, and using the number conservation equation~\eqref{phieq_brown}, we have
\begin{align}
    U^\nu \nabla^\mu T_{\mu\nu} &=  \frac{f_L}{2\kappa}\frac{\partial \rho}{\partial n}\nabla^\mu \bigl(nU_\mu\bigr) = 0 \,.
\end{align}
As usual for this approach, the conservation along the fluid flow is preserved~\cite{Brown_1993}.

On the other hand, the contraction perpendicular to the direction of flow of~\eqref{Brown_field_cons_3} directly gives
\begin{align}
\label{brown_cons_perpen_field}
    h^\nu_\sigma \nabla^\mu T_{\mu\nu}  
    &=h^\nu_\sigma \Bigl( \frac{f_L}{2\kappa} \frac{\partial \rho}{\partial s}\nabla_\nu s  - n \nabla^\mu\bigl(\frac{f_L}{2\kappa}\frac{\partial \rho}{\partial n}h_{\mu\nu} \bigr)  \Bigr) \neq 0 \,,
\end{align}
where we used $h^\nu_\sigma U_\nu = 0$ such that the second term on the right-hand side of~\eqref{Brown_field_cons_3} vanishes. The remaining term now does not vanish and thus the energy-momentum tensor is not conserved in the perpendicular direction. This result is in contrast with the approach of Schutz, where the energy-momentum tensor was conserved in the perpendicular direction but not in the fluid flow  direction. Non-minimal entropy couplings using the Brown approach were studied in the interacting darks sector models of \cite{Jensko:2026taf}, giving a similar perpendicular source term to the energy-momentum conservation equation.

\subsubsection{Conservation equations from the fluid equation}
\label{conservation equations from the fluid equation}

Now starting from~\eqref{Fluid_T}, where in this approach the energy density $\rho(n,s)$ is the fundamental quantity and the pressure $p$ is derived quantity, and taking the covariant derivative, gives
\begin{align}
\label{convs_T_fluid}
    \nabla^\mu T_{\mu\nu} = \nabla^\mu \bigl(\rho U_\mu U_\nu \bigr) + \nabla^\mu \bigl(ph_{\mu\nu} \bigr) \,.
\end{align}
Substituting the definition of the pressure~\eqref{pressuredef} and contracting along the fluid flow leads to
\begin{align}
    U^\nu \nabla^\mu T_{\mu\nu} &= -\nabla^\mu \bigl(\rho U_\mu \bigr) + n\frac{\partial \rho}{\partial n}U^\nu\nabla^\mu h_{\mu\nu} - \rho U^\nu \nabla^\mu h_{\mu\nu} \notag \\
    &= -\rho\nabla^\mu U_\mu-U_\mu\nabla^\mu\rho - n\frac{\partial \rho}{\partial n}\nabla^\mu U_\mu+\rho\nabla^\mu U_\mu \notag \\
    &= -\frac{\partial \rho}{\partial n}U_\mu\nabla^\mu n-\frac{\partial \rho}{\partial s}U_\mu\nabla^\mu s - n\frac{\partial \rho}{\partial n}\nabla^\mu U_\mu \notag \\
    &= - \frac{\partial \rho}{\partial n}\nabla^\mu\bigl(nU_\mu \bigr)-\frac{\partial \rho}{\partial s}U_\mu\nabla^\mu s = 0 \,.
\end{align}
In the last step we used equations~\eqref{phieq_brown} and~\eqref{thetaeq_brown}. Now contracting~\eqref{convs_T_fluid} with $h^\nu_\sigma$ for the direction perpendicular to the flow, after some simple  manipulations we find
\begin{align}
    \label{Brown_mid_perpen_cons_fluid}
    h^\nu_\sigma \nabla^\mu T_{\mu\nu} = 2nU^\nu \nabla_{[\nu}\bigl( \upmu U_{\sigma]} \bigr) - \frac{\partial \rho}{\partial s}\nabla_\sigma s\,,
\end{align}
where we have used the definition~\eqref{chemical_potential} for the chemical potential. We also have that~\eqref{Jeq_Brown_effective} can be antisymmetrized to give
\begin{align}
    2U^\nu\nabla_{[\nu}\bigl( \upmu_{\rm eff}U_{\sigma]}\bigr) = U^\nu \theta_{,\nu}\nabla_\sigma s \,.
\end{align}
To arrive at this result we also used
\begin{align}
    U^\nu \nabla_{[\nu} \bigl(\varphi_{,\sigma]}\bigr) = 0 \,, \quad  U^\nu \nabla_{[\nu} \bigl(s\theta_{,\sigma]}\bigr) = -U^\nu \theta_{,\nu} \nabla_\sigma s\,, \quad  U^\nu \nabla_{[\nu} \bigl(\beta_A\alpha^A_{,\sigma]}\bigr) = 0\,.
\end{align}

Using the definition of the effective chemical potential~\eqref{effective_chemical_potential} and using~\eqref{seq_brown} gives
\begin{align}
    2nU^\nu\nabla_{[\nu}\bigl( \upmu U_{\sigma]}\bigr) = \bigl( \frac{f_L}{2\kappa}+1 \bigr)\frac{\partial \rho}{\partial s}\nabla_\sigma s - 2nU^\nu\nabla_{[\nu}\bigl( \frac{f_L}{2\kappa}\upmu U_{\sigma]}\bigr) \,.
\end{align}
Substituting this result back in~\eqref{Brown_mid_perpen_cons_fluid} gives
\begin{align}
     h^\nu_\sigma \nabla^\mu T_{\mu\nu} =  \frac{f_L}{2\kappa}\frac{\partial \rho}{\partial s}\nabla_\sigma s - 2nU^\nu\nabla_{[\nu}\bigl( \frac{f_L}{2\kappa}\upmu U_{\sigma]}\bigr)  \,.
\end{align}
Finally, this simplifies to
\begin{align}
     h^\nu_\sigma \nabla^\mu T_{\mu\nu} = h^\nu_\sigma \Bigl(\frac{f_L}{2\kappa}\frac{\partial \rho}{\partial s}\nabla_\nu s - n\nabla_\mu \bigl( h^\mu_\nu \frac{f_L}{2\kappa}\frac{\partial \rho}{\partial n}\bigr) \Bigr) \neq 0 \,,
\end{align}
which matches the result in~\eqref{brown_cons_perpen_field}. (For a detailed derivation, see Section A.4 in \cite{Boehmer:2025afy}.) This demonstrates the consistency in deriving  
the conservation equations, starting both from the field equations and from the relativistic fluid equations. It also shows the clear distinction between the Brown and Schutz conservation equations, which lead to different physical outcomes.

\subsection{Effective equations and interactions} 
\label{Brown_dark_intr}
 
We now investigate how the effective contribution to the field equations~\eqref{Brown_field_eqn} in this approach can shed light on dark sector interactions by considering the same coupling function $f(R,L_{\rm m}) = R+\bar{f}(R,L_{\rm m})$ used earlier in Section~\ref{Schutz_dark_intr}. The Einstein tensor from~\eqref{Brown_field_eqn} can be written as
\begin{align}
\label{einstein_brown_dark}
    G_{\mu\nu} =  \kappa T_{\mu\nu} + \frac{1}{2} \bar{f}_L n\frac{\partial \rho}{\partial n} \bigl(g_{\mu\nu}+U_\mu U_\nu \bigr) + \frac{1}{2} \bar{f} g_{\mu\nu} - \bar{f}_R R_{\mu\nu} - \big(g_{\mu\nu}\Box - \nabla_\mu\nabla_\nu\big) \bar{f}_R\,.
\end{align}
As in Section~\ref{Schutz_dark_intr}, we can make an effective decomposition of the modified field equations.
Since the identification of dark sector terms is arbitrary, we use a different identification this time. An immediate choice of identification is the following term,
\begin{align}
    T_{\mu\nu}^{\rm DM} :=   \frac{1}{2\kappa} \bar{f}_L n\frac{\partial \rho}{\partial n} U_{\mu} U_{\nu} \,,
\end{align}
which resembles a pressureless cold dark matter contribution. 
One can also identify a piece which resembles the conventional form of the cosmological constant term $\Lambda g_{\mu\nu}$, 
\begin{align}
    T_{\mu\nu}^{\Lambda} :=  \frac{1}{2\kappa} \Bigl(\bar{f} + \bar{f}_L n\frac{\partial \rho}{\partial n}\Bigr)g_{\mu\nu} \,.
\end{align}
Finally, the remaining components in~\eqref{einstein_brown_dark} are related to the modified $\bar{f}_{R}$ terms,
\begin{align}
    T_{\mu\nu}^{\bar{f}} := - \frac{1}{\kappa}\bigl(R_{\mu\nu} + g_{\mu\nu}\Box - \nabla_\mu\nabla_\nu\bigr) \bar{f}_R\,,
\end{align}
which are best interpreted as further contributions to the dark energy sector, i.e., $T_{\mu\nu}^{\rm DE} = T_{\mu\nu}^{\Lambda} + T_{\mu\nu}^{\bar{f}}$.
These definitions also resonate with those used in~\cite{Boehmer:2025afy} to allow the comparison of the results. The Einstein tensor can now be written as
\begin{align}
    G_{\mu\nu} =  \kappa \Bigl(T_{\mu\nu} + T_{\mu\nu}^{\rm DM} + T_{\mu\nu}^{\Lambda} + T_{\mu\nu}^{\bar{f}}\Bigr) \,.
\end{align}
Taking the covariant derivative leads to the following conservation relation:
\begin{align}
    \nabla^\mu \Bigl(T_{\mu\nu} + T_{\mu\nu}^{\rm DM} + T_{\mu\nu}^{\Lambda} \Bigr) =
    -\nabla^\mu T_{\mu\nu}^{\bar{f}} =\frac{1}{2\kappa}\bar{f}_{R}\nabla_\nu R =: Q_\nu \,,
\end{align}
where $Q_\nu$ is a dark interaction term. The result coincides with~\cite{Boehmer:2025afy} where the separation between the dark matter term and dark energy terms is not obvious. Such difficulties have been addressed in the literature (see~\cite{Bansal:2024bbb, CarrilloGonzalez:2017cll}).
We state again that the identifications of the terms is arbitrary and based on the features expected for dark matter and dark energy. 

\section{Similarities and differences}
\label{Implications}

\subsection{Enthalpy}

We begin by recalling the definition of enthalpy $H=\rho+p$, which using the relation in~\eqref{chemical_potential} gives $H=\upmu n$. Interestingly, this quantity can be defined using the effective quantities introduced in either approach and we find that both coincide, namely
\begin{align}
    H_{\rm eff}^{\rm Schutz} = \upmu n_{\rm eff} = \upmu \Bigl(1+\frac{f_L}{2\kappa}\Bigr)n = 
     \Bigl(1+\frac{f_L}{2\kappa}\Bigr)\upmu n = \upmu_{\rm eff} n = H_{\rm eff}^{\rm Brown} \,. 
\end{align}
This can be interpreted as follows: in both approaches the same amount of energy is introduced into the system due to non-minimal couplings. This energy represents the additional coupling between the geometry and the matter. However, it does so in two distinct ways, either via an effective energy density or an effective pressure. Consequently, the energy transfer takes place either along the direction of the flow or perpendicular to the flow. 

In Appendix~\ref{appendixA} we provide Table~\ref{tabappendix} which contains a side-by-side comparison of all quantities.

\subsection{Conservation equations}

One can immediately note that the effective energy-momentum tensors in~\eqref{effective_T_schutz} and~\eqref{effective_T_Brown}, can be written as
\begin{alignat}{2}
    T_{\mu \nu}^{\textrm{eff}} &= T_{\mu\nu}+\frac{1}{2\kappa}f_L (\rho+p)U_\mu U_\nu  \qquad &\textrm{Schutz's approach} \,, \label{eff_schutz_T2}\\
    T_{\mu \nu}^{\textrm{eff}} &= T_{\mu\nu}+\frac{1}{2\kappa}f_L (\rho+p) h_{\mu\nu}  \label{eff_brown_T2} \qquad &\textrm{Brown's approach} \,.
\end{alignat}
Both of these results are consistent with the field equations originally derived in~\cite{Harko:2010mv}, where the effective-energy-momentum tensor can be identified with the general form
\begin{align}
    \label{Harko_eff_T}
    T_{\mu \nu}^{\textrm{eff}} &= T_{\mu\nu}+\frac{1}{2\kappa}f_L (T_{\mu\nu}-L_{\rm m} g_{\mu\nu}) \,.
\end{align}
It is important to see here that changing the matter Lagrangian in~\eqref{Harko_eff_T} can change the underlying physical theory even if different Lagrangians describe the same physical system. Our results in~\eqref{eff_schutz_T2} and~\eqref{eff_brown_T2} clearly show that, while the two approaches are analogous in form, they lead to different field equations. The contribution of the effective energy-momentum tensor is Schutz's approach is in the fluid flow direction, while in Brown's approach it is in the perpendicular direction. Because of this, one can define an effective energy density in Schutz's approach and an effective pressure in Brown's.

The general expression for the covariant derivative of the energy-momentum tensor given in~\cite{Harko:2010mv} provides another direct way to show that distinct theories can emerge,
\begin{align}
    \label{General_cons_eqn}
     \nabla^\mu T_{\mu\nu} = \big(L_{\rm m} g_{\mu\nu} - T_{\mu\nu} \big)\nabla^\mu \log\Big(1+\frac{f_L}{2\kappa}\Big ) \,.
\end{align}
Using the appropriate fluids descriptions, we have 
\begin{alignat}{2}
    \nabla^\mu T_{\mu\nu} &= - \upmu \frac{\partial p}{\partial \upmu}U_\mu U_\nu\nabla^\mu \log \Big(1+\frac{f_L}{2\kappa}\Big) = -U_\nu Q &\qquad &\textrm{Schutz approach} \,,  \label{Schutz_cons_1} \\
    \nabla^\mu T_{\mu\nu} &= -n \frac{\partial \rho}{\partial n}h_{\mu\nu} \nabla^\mu \log \Big(1+\frac{f_L}{2\kappa}\Big) = -h_{\mu\nu}W^\mu &\qquad &\textrm{Brown approach}   \,, \label{Brown_cons_1}
\end{alignat}
where we introduced the following quantities
\begin{align}
    Q &:= \upmu \frac{\partial p}{\partial \upmu} U_\mu \nabla^\mu \log \Big(1+\frac{f_L}{2\kappa}\Big)\,, \\
    W^\mu &:= n \frac{\partial\rho}{\partial n} \nabla^\mu \log \Big(1+\frac{f_L}{2\kappa}\Big) \,. 
\end{align}
This matches the results in Sections~\ref{sec_Schutz_conservtion_eqs} and~\ref{Brown_conservation_equations}. We can immediately note that for a perfect fluid the energy-momentum tensor is not conserved in general. Schutz's approach mediates a transfer of energy alone, while Brown's approach describes momentum transfer, emphasising once more the distinct nature of the theories that emerge using the different approaches. Our results agree with recent literature~\cite{Lobo:2025eex,BarrosoVarela:2025mro} where it was also discussed that models with non-minimal couplings can lead to distinct physical results.

\subsection{Comparison with $f(R,T)$ gravity}

In~\cite{Boehmer:2025afy}, we established the correct form of the field equations and fluid equations for relativistic perfect fluids in $f(R,T)$ gravity. The non-minimal couplings of $f(R,T)$ and $f(R,L_m)$ models lead to an effective energy-momentum tensor, see Table~\ref{table2}.

\begin{table}[!htb]
    \centering
    \begin{tabular}{c|cc}
        Model & Schutz's approach & Brown's approach \\
        \hline
        $f(R,L_{\rm m})$ & 
        ${\displaystyle T_{\mu\nu}^{\rm eff} = T_{\mu\nu}+\frac{1}{2\kappa}f_L\upmu\frac{\partial p}{\partial \upmu} U_\mu U_\nu}$ &
        ${\displaystyle T_{\mu \nu}^{\textrm{eff}} = T_{\mu\nu}-\frac{1}{2\kappa}f_L n\frac{\partial (-\rho)}{\partial n} h_{\mu\nu}}$ 
        \\[2ex]
        $f(R,T)$ &
        ${\displaystyle T_{\mu\nu}^{\rm eff} = T_{\mu\nu} + \frac{1}{2\kappa}f_T \upmu\frac{\partial T}{\partial \upmu} U_\mu U_\nu}$ &
        ${\displaystyle T^{\rm eff}_{\mu\nu} =T_{\mu\nu} - \frac{1}{2\kappa} f_T n\frac{\partial T}{\partial n}h_{\mu\nu}}$
    \end{tabular}
    \caption{Energy-momentum tensors of $f(R,L_{\rm m})$ and $f(R,T)$ models.}
    \label{table2}
\end{table}
Formally the relationship is simply replacing the matter Lagrangian with the trace or vice versa, $L_m \leftrightarrow T$, and the expressions in one setting correspond to the other. This approach can be applied to the field equations directly too. However, the same cannot be said for the fluid equations which display a significantly different behaviour.  

The difference in behaviour can be clearly seen in the effective thermodynamic quantities. The $f(R,L_{\rm m})$ model in Schutz approach defined an effective particle-number density $n_{\rm eff}=(1+\frac{f_L}{2\kappa})\partial p/\partial \upmu$. In the case of $f(R,T)$ models in~\cite{Boehmer:2025afy}, the modified particle-number density is 
\begin{align}
    n_{\rm eff} = \frac{\partial p}{\partial \upmu}+\frac{f_T}{2\kappa}\frac{\partial T}{\partial \upmu} \qquad {\rm for} \ f(R,T) \  \rm models \,.
\end{align}
This is not a mere perturbation to the number density $n=\partial p/\partial \upmu$, for $\partial T/\partial\upmu$ is a term that is given by second order derivatives of $p$ with respect to $\upmu$. This second order derivative arises from the coupling function $f(R,T)$ which does not factor with the matter action given by $L_{\rm m}$. For example, in the entropy variation~\eqref{delta_s_schutz1}, the factor $(1+f_L/2\kappa)$ arises in both $\partial p/\partial \upmu$ and $\partial p/\partial s$ terms and thus factors out. This is not the case in $f(R,T)$ gravity and one cannot find a common factor which leads to an effective temperature $\mathcal{T}_{\rm eff}$.  

This applies to all thermodynamic quantities in both approaches and thus care should be taken with these quantities across the two models. In the case of Brown's approach, this effect manifests in how the Helmholtz free energy is defined, where we have
\begin{align}
    f(R,L_{\rm m})&: \quad  U^\mu\varphi_{,\mu} = \upmu_{\rm eff} - s\mathcal{T}_{\rm eff} =  \bigl(1+\frac{f_L}{2\kappa}\bigr)\bigl(\upmu-s\mathcal{T} \bigr) \,,\\
    f(R,T)&: \quad U^\mu\varphi_{,\mu} =\upmu_{\rm eff} - s\mathcal{T}_{\rm eff} = \upmu - \frac{f_T}{2\kappa}\frac{\partial T}{\partial n}-s\bigl(\mathcal{T}-\frac{1}{n}\frac{f_T}{2\kappa}\frac{\partial T}{\partial s} \bigr) \,.
\end{align}
While we get an effective Helmholtz free energy of some sort in $f(R,T)$ gravity, it is much more complicated and does not give the same elegant result as in~\eqref{effective_Helmholtz_brown}. 

We see thus that the difference between $f(R,L_{\rm m})$ and $f(R,T)$ models is not just in the nature of the effective term, but also in whether an effective thermodynamic quantity arises or not.
\subsection{Trivial and non-trivial models}
\label{Trivial and non-trivial models section}
Let us assume a separable function of the form $f=f^{(1)}(R) + f^{(2)}(L_m)$ in either approach, then the respective field equations are given by
\begin{alignat}{2}
\textrm{Schutz}&: \notag \\
    &f^{(1)}_R R_{\mu\nu} + \big(g_{\mu\nu}\square - \nabla_\mu \nabla_\nu \big)f^{(1)}_R-\frac{1}{2}f^{(1)}g_{\mu\nu} =\kappa T_{\mu\nu}+\frac{1}{2}f^{(2)}_L\upmu\frac{\partial p}{\partial \upmu} U_\mu U_\nu + 
    \frac{1}{2}f^{(2)}g_{\mu\nu}\,,
    %first equation is Schutz
    \\
    \textrm{Brown}&: \notag \\
    &f^{(1)}_R R_{\mu\nu} + \big(g_{\mu\nu}\square - \nabla_\mu \nabla_\nu \big)f^{(1)}_R   - \frac{1}{2}f^{(1)}g_{\mu\nu} = 
    \kappa T_{\mu\nu}+\frac{1}{2}f^{(2)}_L n\frac{\partial \rho}{\partial n} h_{\mu\nu} + 
    \frac{1}{2}f^{(2)}g_{\mu\nu} \,.
    %second equation os Brown
\end{alignat}
We note that the left-hand sides are equivalent to the $f(R)$ field equations, while the right-hand sides can simply be seen as some induced energy-momentum tensor we can call $\widetilde{T}_{\mu\nu}$. This is nothing but $f(R)$ gravity with an usually constructed matter source, similar to the corresponding result in $f(R,T)$ gravity, see~\cite{Boehmer:2025afy}. No gravitational experiment could ever distinguish this model from $f(R)$ gravity with a different effective matter source.

The condition that defines a non-trivial model, with genuine matter-curvature couplings, can be read off from the conservation~\eqref{General_cons_eqn}. Minimal matter-curvature couplings of the form $f(R,L_{\rm m})=f^{(1)}(R)+f^{(2)}(L_{\rm m})$, will produce a right-hand side of~\eqref{General_cons_eqn} that is independent of geometry. Thus all such models are nothing but $f(R)$ gravity with effective thermodynamic terms. Consequently, any non-trivial model, where matter and geometry couple meaningfully, is defined by $f_{RL} \neq 0$. This conclusion is expected because of the previous results in~\cite{Boehmer:2025afy}.

\section{Conclusions}

We have considered $f(R,L_m)$ curvature-matter coupling models for relativistic perfect fluids using two approaches: Schutz's formulation using velocity potentials and Brown's formulation using Lagrange multipliers. Each approach treats a different thermodynamic quantity as fundamental, leading to different field equations and effective energy-momentum tensors. These differences are reflected in the modified thermodynamic quantities and the conservation equations. In Schutz's formulation, the non-minimal coupling induces an effective particle number density, whereas in Brown's formulation it induces an effective chemical potential, temperature, and Helmholtz free energy (see Table~\ref{tabappendix}). A key physical result is that the divergence of the energy-momentum tensor $T_{\mu\nu}$ is parallel to the fluid flow in Schutz's approach and perpendicular to it in Brown's approach. Consequently, the former approach is better suited to describe energy transfer while the latter describes momentum transfer.
Thus, care must be taken when choosing which matter Lagrangian to use, as the physical implications are distinct.

A comparison with our previous result in~\cite{Boehmer:2025afy} indicate that our result is also consistent with that for $f(R,T)$ gravity. One can see from Table~\ref{table2} that the effective term in each model is controlled by the derivative of the non-minimal coupling function $f$ with respect to the relevant thermodynamic quantity. The key fundamental difference between the two models is that the trace of the energy-momentum tensor in the case of $f(R,T)$ gravity includes second-order variations in the Lagrangian which distinguishes it from $f(R,L_{\rm m})$ models, which only contains first variations. If one were to incorrectly neglect the second-order variations in $f(R,T)$ gravity, the equations would reduce to the $f(R,L_{\rm m})$ model.

The on-shell analysis shows that the usual minimally coupled relation between the pressure and energy density Lagrangians is modified by the non-minimal coupling, as shown in Sections~\ref{On-shell Argument Schutz} and~\ref{On-shell Argument Brown}. In Schutz's formulation, the matter action written with $p$ is on-shell equivalent to the one written with $-\rho_{\rm eff}$, while in Brown's formulation the matter action written with $-\rho$ is on-shell equivalent to the one written with $p_{\rm eff}$. These equivalences hold on shell at the level of the matter action, but they do not extend to the $f(R,L_{\rm m})$ coupling itself. Thus, the Schutz and Brown choices remain generically inequivalent, even on shell, for non-minimal curvature-matter couplings.

In order to describe the dark sector, one may split the effective source terms into dark matter and dark energy components, as discussed in Section~\ref{Schutz_dark_intr} and Section~\ref{Brown_dark_intr}. This leads to an effective interaction between the two components, although the precise split depends on the chosen fluid formulation. In particular, the different projected directions of the additional non-minimal coupling terms in the Schutz and Brown field equations motivate different effective decompositions of the dark sector. However, the separation into distinct dark sector contributions is arbitrary and should not be over-interpreted. More generally, the choice of which terms to label as dark matter or dark energy will influence whether the interactions are interpreted as energy transfer or momentum transfer, as discussed in~\cite{Jensko:2026taf}. These points should be kept in mind, especially in relation to interacting dark sector models, where the interpretation of the interactions is less unique than it may initially appear.

To physically distinguish between the two approaches, one must focus not on the Lagrangian alone but the resulting energy-momentum tensors. For instance, the cosmological signatures of non-minimally coupled models in the Schutz and Brown formalisms will lead to distinct results, and this could be used to assess which constructions are phenomenologically viable. Nonetheless, when studying $f(R,L_{\rm m})$ coupling models, one should be careful to work with models that can actually produce new physics. This means working with non-trivial $f(R,L_{\rm m})$ models in which $f_{RL}\neq 0$. Any other model with trivial coupling is gravitationally indistinguishable from $f(R)$ gravity with a redefined effective matter sector.

\subsection*{Acknowledgments}
EJ is supported by the Engineering and Physical Sciences Research Council (EPSRC) [EP/W524335/1, UKRI3030]. EA acknowledges the funding provided by Kuwait University through its graduate student scholarship programme. 

\appendix
\section{Table of key quantities in both formalisms}
\label{appendixA}

\begin{table}[h]
    \centering
    \begin{tabular}{c|cc}
        Approach & Schutz & Brown  \\
        \hline
        Lagrangian  & $L_{\rm m}=p(\upmu,s)$ & $L_{\rm m}=-\rho(n,s)$ \\
        constraint & $\mathcal{C}_{\rm m} = 0$ & $\mathcal{C}_{\rm m} = J^\mu(\varphi_{,\mu}+s\theta_{,\mu}+\beta_A\alpha^A_{,\mu})$ \\
        derived quantity  & ${\displaystyle \rho=\upmu\frac{\partial p}{\partial\upmu}-p}$ & 
        ${\displaystyle p=n\frac{\partial \rho}{\partial n}-\rho}$ \\
        dependent variable & $\upmu = \sqrt{-V_\nu V^\nu}$ & $n = \sqrt{-J_\nu J^\nu }/\sqrt{-g}$ \\
        independent variables &  $\{ g^{\mu\nu}, \alpha, \beta, \phi, \theta, s \}$ &
        $\{ g^{\mu\nu}, J^\mu, \alpha^A, \beta_A, \varphi, \theta, s\}$ \\
        effective quantity & ${\displaystyle \rho_{\rm eff} = \rho+ \frac{f_L}{2\kappa}\upmu\frac{\partial p}{\partial \upmu}}$ & 
        ${\displaystyle p_{\rm eff} = p+\frac{f_L}{2\kappa}n\frac{\partial \rho}{\partial n}}$ \\
        on-shell Lagrangian & $p=-\rho_{\rm eff}$ & $-\rho=p_{\rm eff}$ \\
        particle number density & $n_{\rm eff} = \bigl(1+\frac{f_L}{2\kappa}\bigr)n$ & $n$ \\
        chemical potential & $\upmu$ & ${\displaystyle \upmu_{\rm eff} = \bigl(1+\frac{f_L}{2\kappa}\bigr)\upmu}$ \\
        temperature & $\mathcal{T}$ & ${\displaystyle \mathcal{T}_{\rm eff} = \bigl(1+\frac{f_L}{2\kappa}\bigr)\mathcal{T}}$ \\
        enthalpy & $H_{\rm eff} = \upmu n_{\rm eff}$ & $H_{\rm eff} = \upmu_{\rm eff} n$ \\ 
        free energy & $\mathcal{F}=\upmu-\mathcal{T}s$ & ${\displaystyle \mathcal{F}_{\rm eff}= \bigl(1+\frac{f_L}{2\kappa}\bigr)\mathcal{F}}$ \\
        conservation Eqs. & $\nabla^\mu T_{\mu\nu} = -U_\nu Q$ & $\nabla^\mu T_{\mu\nu} = -h_{\mu\nu} W^\mu$
    \end{tabular}
    \caption{Summary of the resulting effective terms and on-shell expression}
    \label{tabappendix}
\end{table}

\addcontentsline{toc}{section}{References}
\bibliographystyle{jhepmodstyle}
\bibliography{bib}

@article{Clifton:2011jh,
    author = "Clifton, Timothy and Ferreira, Pedro G. and Padilla, Antonio and Skordis, Constantinos",
    title = "{Modified Gravity and Cosmology}",
    eprint = "1106.2476",
    archivePrefix = "arXiv",
    primaryClass = "astro-ph.CO",
    doi = "10.1016/j.physrep.2012.01.001",
    journal = "Phys. Rept.",
    volume = "513",
    pages = "1--189",
    year = "2012"
}

@article{Boehmer:2025afy,
    author = "B{\"o}ehmer, Christian G. and Al-Nasrallah, Eissa",
    title = "{Consistent energy-momentum trace couplings of fluids}",
    eprint = "2509.24843",
    archivePrefix = "arXiv",
    primaryClass = "gr-qc",
    doi = "10.1103/6wrh-lg9s",
    journal = "Phys. Rev. D",
    volume = "112",
    number = "12",
    pages = "124073",
    year = "2025"
}

@article{Jensko:2026taf,
    author = "Jensko, Erik and Teixeira, Elsa M. and Poulin, Vivian",
    title = "{Interacting dark sector from intrinsic entropy couplings}",
    eprint = "2603.10622",
    archivePrefix = "arXiv",
    primaryClass = "astro-ph.CO",
    month = "3",
    year = "2026"
}

@article{Wang:2024vmw,
    author = "Wang, B. and Abdalla, E. and Atrio-Barandela, F. and Pav{\'o}n, D.",
    title = "{Further understanding the interaction between dark energy and dark matter: current status and future directions}",
    eprint = "2402.00819",
    archivePrefix = "arXiv",
    primaryClass = "astro-ph.CO",
    doi = "10.1088/1361-6633/ad2527",
    journal = "Rept. Prog. Phys.",
    volume = "87",
    number = "3",
    pages = "036901",
    year = "2024"
}

@article{Wang:2016lxa,
    author = "Wang, B. and Abdalla, E. and Atrio-Barandela, F. and Pavon, D.",
    title = "{Dark Matter and Dark Energy Interactions: Theoretical Challenges, Cosmological Implications and Observational Signatures}",
    eprint = "1603.08299",
    archivePrefix = "arXiv",
    primaryClass = "astro-ph.CO",
    doi = "10.1088/0034-4885/79/9/096901",
    journal = "Rept. Prog. Phys.",
    volume = "79",
    number = "9",
    pages = "096901",
    year = "2016"
}

@article{Bertolami:2008ab,
    author = "Bertolami, Orfeu and Lobo, Francisco S. N. and Paramos, Jorge",
    title = "{Non-minimum coupling of perfect fluids to curvature}",
    eprint = "0806.4434",
    archivePrefix = "arXiv",
    primaryClass = "gr-qc",
    doi = "10.1103/PhysRevD.78.064036",
    journal = "Phys. Rev. D",
    volume = "78",
    pages = "064036",
    year = "2008"
}

@article{Li:2026xaz,
    author = "Li, Tian-Nuo and Giar{\`e}, William and Du, Guo-Hong and Li, Yun-He and Di Valentino, Eleonora and Zhang, Jing-Fei and Zhang, Xin",
    title = "{Strong Evidence for Dark Sector Interactions}",
    eprint = "2601.07361",
    archivePrefix = "arXiv",
    primaryClass = "astro-ph.CO",
    month = "1",
    year = "2026"
}

@article{Pourtsidou:2013nha,
    author = "Pourtsidou, A. and Skordis, C. and Copeland, E. J.",
    title = "{Models of dark matter coupled to dark energy}",
    eprint = "1307.0458",
    archivePrefix = "arXiv",
    primaryClass = "astro-ph.CO",
    doi = "10.1103/PhysRevD.88.083505",
    journal = "Phys. Rev. D",
    volume = "88",
    number = "8",
    pages = "083505",
    year = "2013"
}

@article{Amendola:1999er,
    author = "Amendola, Luca",
    title = "{Coupled quintessence}",
    eprint = "astro-ph/9908023",
    archivePrefix = "arXiv",
    doi = "10.1103/PhysRevD.62.043511",
    journal = "Phys. Rev. D",
    volume = "62",
    pages = "043511",
    year = "2000"
}

@article{Sotiriou:2008rp,
    author = "Sotiriou, Thomas P. and Faraoni, Valerio",
    title = "{f(R) Theories Of Gravity}",
    eprint = "0805.1726",
    archivePrefix = "arXiv",
    primaryClass = "gr-qc",
    doi = "10.1103/RevModPhys.82.451",
    journal = "Rev. Mod. Phys.",
    volume = "82",
    pages = "451--497",
    year = "2010"
}

@article{DeFelice:2010aj,
    author = "De Felice, Antonio and Tsujikawa, Shinji",
    title = "{f(R) theories}",
    eprint = "1002.4928",
    archivePrefix = "arXiv",
    primaryClass = "gr-qc",
    doi = "10.12942/lrr-2010-3",
    journal = "Living Rev. Rel.",
    volume = "13",
    pages = "3",
    year = "2010"
}

@article{CosmoVerseNetwork:2025alb,
    author = "Di Valentino, Eleonora and others",
    collaboration = "CosmoVerse Network",
    title = "{The CosmoVerse White Paper: Addressing observational tensions in cosmology with systematics and fundamental physics}",
    eprint = "2504.01669",
    archivePrefix = "arXiv",
    primaryClass = "astro-ph.CO",
    doi = "10.1016/j.dark.2025.101965",
    journal = "Phys. Dark Univ.",
    volume = "49",
    pages = "101965",
    year = "2025"
}

@article{Copeland:2006wr,
    author = "Copeland, Edmund J. and Sami, M. and Tsujikawa, Shinji",
    title = "{Dynamics of dark energy}",
    eprint = "hep-th/0603057",
    archivePrefix = "arXiv",
    doi = "10.1142/S021827180600942X",
    journal = "Int. J. Mod. Phys. D",
    volume = "15",
    pages = "1753--1936",
    year = "2006"
}

@article{Harko:2010mv,
    author = "Harko, Tiberiu and Lobo, Francisco S. N.",
    title = "{f(R,$L_{m}$) gravity}",
    eprint = "1008.4193",
    archivePrefix = "arXiv",
    primaryClass = "gr-qc",
    doi = "10.1140/epjc/s10052-010-1467-3",
    journal = "Eur. Phys. J. C",
    volume = "70",
    pages = "373--379",
    year = "2010"
}

@article{Schutz:1970,
  title = {Perfect Fluids in General Relativity: Velocity Potentials and a Variational Principle},
  author = {Schutz, Bernard F.},
  journal = {Phys. Rev. D},
  volume = {2},
  issue = {12},
  pages = {2762--2773},
  numpages = {0},
  year = {1970},
  month = {Dec},
  publisher = {American Physical Society},
  doi = {10.1103/PhysRevD.2.2762},
  url = {https://link.aps.org/doi/10.1103/PhysRevD.2.2762}
}

@article{Brown_1993,
	title        = {Action functionals for relativistic perfect fluids},
	author       = {J D Brown},
	year         = 1993,
	month        = 8,
	journal      = {Classical and Quantum Gravity},
	publisher    = {{IOP} Publishing},
	volume       = 10,
	number       = 8,
	pages        = {1579--1606},
	doi          = {10.1088/0264-9381/10/8/017},
	url          = {https://doi.org/10.1088\%2F0264-9381\%2F10\%2F8\%2F017}
}

@article{Iosifidis:2024ksa,
    author = "Iosifidis, Damianos and Jensko, Erik and Koivisto, Tomi S.",
    title = "{Relativistic interacting fluids in cosmology}",
    eprint = "2406.01412",
    archivePrefix = "arXiv",
    primaryClass = "gr-qc",
    doi = "10.1088/1475-7516/2024/11/043",
    journal = "JCAP",
    volume = "11",
    pages = "043",
    year = "2024"
}

@book{Misner:1973prb,
    author = "Misner, Charles W. and Thorne, K. S. and Wheeler, J. A.",
    title = "{Gravitation}",
    isbn = "978-0-7167-0344-0, 978-0-691-17779-3",
    publisher = "W. H. Freeman",
    address = "San Francisco",
    year = "1973"
}

@book{CANTATA:2021asi,
    author = "Akrami, Yashar and others",
    editor = "Saridakis, Emmanuel N. and Lazkoz, Ruth and Salzano, Vincenzo and Vargas Moniz, Paulo and Capozziello, Salvatore and Beltr{\'a}n Jim{\'e}nez, Jose and De Laurentis, Mariafelicia and Olmo, Gonzalo J.",
    collaboration = "CANTATA",
    title = "{Modified Gravity and Cosmology. An Update by the CANTATA Network}",
    eprint = "2105.12582",
    archivePrefix = "arXiv",
    primaryClass = "gr-qc",
    doi = "10.1007/978-3-030-83715-0",
    isbn = "978-3-030-83714-3, 978-3-030-83717-4, 978-3-030-83715-0",
    publisher = "Springer",
    year = "2021"
}

@article{Bertolami:2007gv,
    author = "Bertolami, Orfeu and B{\"o}hmer, Christian G. and Harko, Tiberiu and Lobo, Francisco S. N.",
    title = "{Extra force in f(R) modified theories of gravity}",
    eprint = "0704.1733",
    archivePrefix = "arXiv",
    primaryClass = "gr-qc",
    doi = "10.1103/PhysRevD.75.104016",
    journal = "Phys. Rev. D",
    volume = "75",
    pages = "104016",
    year = "2007"
}

@article{NOJIRI2004137,
title = {Gravity assisted dark energy dominance and cosmic acceleration},
journal = {Physics Letters B},
volume = {599},
number = {3},
pages = {137-142},
year = {2004},
issn = {0370-2693},
doi = {https://doi.org/10.1016/j.physletb.2004.08.045},
url = {https://www.sciencedirect.com/science/article/pii/S0370269304012201},
author = {Shin'ichi Nojiri and Sergei D. Odintsov}
}

@article{Bertolami_2008,
   title={Nonminimal coupling of perfect fluids to curvature},
   volume={78},
   ISSN={1550-2368},
   url={http://dx.doi.org/10.1103/PhysRevD.78.064036},
   DOI={10.1103/physrevd.78.064036},
   number={6},
   journal={Physical Review D},
   publisher={American Physical Society (APS)},
   author={Bertolami, Orfeu and Lobo, Francisco S. N. and Páramos, Jorge},
   year={2008},
   month=sep }

@article{Bertolami:2008zh,
    author = "Bertolami, Orfeu and Paramos, Jorge and Harko, Tiberiu and Lobo, Francisco S. N.",
    title = "{Non-minimal curvature-matter couplings in modified gravity}",
    eprint = "0811.2876",
    archivePrefix = "arXiv",
    primaryClass = "gr-qc",
    month = "11",
    year = "2008"
}

@article{Bansal:2024bbb,
    author = "Bansal, Pulkit and Johnson, Joseph P. and Shankaranarayanan, S.",
    title = "{Interacting dark sector from Horndeski theories and beyond: Mapping fields and fluids}",
    eprint = "2408.12341",
    archivePrefix = "arXiv",
    primaryClass = "astro-ph.CO",
    doi = "10.1103/PhysRevD.111.024071",
    journal = "Phys. Rev. D",
    volume = "111",
    number = "2",
    pages = "024071",
    year = "2025"
}

@article{CarrilloGonzalez:2017cll,
    author = "Carrillo Gonz{\'a}lez, Mariana and Trodden, Mark",
    title = "{Field Theories and Fluids for an Interacting Dark Sector}",
    eprint = "1705.04737",
    archivePrefix = "arXiv",
    primaryClass = "astro-ph.CO",
    doi = "10.1103/PhysRevD.97.043508",
    journal = "Phys. Rev. D",
    volume = "97",
    number = "4",
    pages = "043508",
    year = "2018",
    note = "[Erratum: Phys.Rev.D 101, 089901 (2020)]"
}

@article{BarrosoVarela:2025mro,
    author = "Barroso Varela, Miguel and Bertolami, Orfeu",
    title = "{Density perturbations in nonminimally coupled gravity: symptoms of Lagrangian density ambiguity}",
    eprint = "2505.10291",
    archivePrefix = "arXiv",
    primaryClass = "gr-qc",
    doi = "10.1088/1475-7516/2026/01/032",
    journal = "JCAP",
    volume = "01",
    pages = "032",
    year = "2026"
}

@article{Lobo:2025eex,
    author = "Lobo, Francisco S. N. and Harko, Tiberiu and Pinto, Miguel A. S.",
    title = "{Modified Gravity with Nonminimal Curvature{\textendash}Matter Couplings: A Framework for Gravitationally Induced Particle Creation}",
    eprint = "2510.24371",
    archivePrefix = "arXiv",
    primaryClass = "gr-qc",
    doi = "10.3390/universe11110356",
    journal = "Universe",
    volume = "11",
    number = "11",
    pages = "356",
    year = "2025"
}

@article{Ballesteros:2016kdx,
    author = "Ballesteros, Guillermo and Comelli, Denis and Pilo, Luigi",
    title = "{Thermodynamics of perfect fluids from scalar field theory}",
    eprint = "1605.05304",
    archivePrefix = "arXiv",
    primaryClass = "hep-th",
    reportNumber = "CERN-TH-2016-098",
    doi = "10.1103/PhysRevD.94.025034",
    journal = "Phys. Rev. D",
    volume = "94",
    number = "2",
    pages = "025034",
    year = "2016"
}

@article{Carter:1987qr,
    author = "Carter, Brandon",
    title = "{Covariant Theory of Conductivity in Ideal Fluid or Solid Media}",
    reportNumber = "NSF-ITP-87-97",
    journal = "Lect. Notes Math.",
    volume = "1385",
    pages = "1--64",
    year = "1989"
}

@article{Andersson:2006nr,
    author = "Andersson, N. and Comer, G. L.",
    title = "{Relativistic fluid dynamics: Physics for many different scales}",
    eprint = "gr-qc/0605010",
    archivePrefix = "arXiv",
    doi = "10.12942/lrr-2007-1",
    journal = "Living Rev. Rel.",
    volume = "10",
    pages = "1",
    year = "2007"
}

@article{Ballesteros:2012kv,
    author = "Ballesteros, Guillermo and Bellazzini, Brando",
    title = "{Effective perfect fluids in cosmology}",
    eprint = "1210.1561",
    archivePrefix = "arXiv",
    primaryClass = "hep-th",
    doi = "10.1088/1475-7516/2013/04/001",
    journal = "JCAP",
    volume = "04",
    pages = "001",
    year = "2013"
}

@article{Dubovsky:2011sj,
    author = "Dubovsky, Sergei and Hui, Lam and Nicolis, Alberto and Son, Dam Thanh",
    title = "{Effective field theory for hydrodynamics: thermodynamics, and the derivative expansion}",
    eprint = "1107.0731",
    archivePrefix = "arXiv",
    primaryClass = "hep-th",
    doi = "10.1103/PhysRevD.85.085029",
    journal = "Phys. Rev. D",
    volume = "85",
    pages = "085029",
    year = "2012"
}

\end{document}